\documentstyle[twocolumn,prl,aps]{revtex}
\input epsf

\begin{document}

\draft
\twocolumn[\hsize\textwidth\columnwidth\hsize\csname@twocolumnfalse\endcsname

\title{Growing spatial correlations of particle displacements in a
simulated liquid on cooling toward the glass transition}

\author{Claudio Donati,$^1$
Sharon C. Glotzer$^1$ and 
Peter H. Poole$^2$}

\address{$^1$ Polymers Division and Center for Theoretical and
Computational Materials Science, NIST, Gaithersburg, Maryland, USA
20899}

\address{$^2$Department of Applied Mathematics,
University of Western Ontario, London, Ontario N6A~5B7, Canada}

\date{Submitted: \today}
\maketitle

\begin{abstract}
We define a correlation function that quantifies the spatial
correlation of single-particle displacements in liquids and amorphous
materials.  We show for an equilibrium liquid that this function is
related to fluctuations in a bulk dynamical variable.  We evaluate
this function using computer simulations of an equilibrium
glass-forming liquid, and show that long range spatial correlations of
displacements emerge and grow on cooling toward the mode coupling
critical temperature.
\end{abstract}

\vskip2pc]
\narrowtext

Liquids cooled toward their glass transition exhibit remarkable
dynamical behavior \cite{ean}.  The initial slowing of transport
processes for liquids at temperatures $T$ well above their glass
transition temperature $T_g$ is described to great extent by the mode
coupling theory (MCT) \cite{gotze,cummins,mctbook}, which predicts
diverging relaxation times at the mode coupling dynamical critical
temperature $T_c$.  The dynamical singularity of MCT occurs without a
diverging or even growing static correlation
length~\cite{vanblaad}.  Yet recent studies show that in the
range of $T$ well described by MCT, simulated glassforming liquids
exhibit spatially heterogeneous
dynamics~\cite{kdppg,ddkppg,dgpkp,hiwatari,harrowell,onuki,ediger}.
In this Letter, we define a correlation function that quantifies the
spatial correlation of particle displacements and evaluate this
function for a simulated Lennard-Jones liquid.  We find that spatial
correlations of displacement arise and become increasingly long ranged
on cooling toward $T_c$.

First, we briefly review the conventional static correlation function
that describes the average microscopic structure of a liquid.  We use
a definition that will readily facilitate an extension to a new
correlation function for particle displacements.  Consider a liquid in
the grand canonical ensemble confined to a volume $V$, consisting of
identical particles, each with no internal degrees of freedom.  Let
the position of each particle $i$ be denoted ${\bf r}_i$.  In
equilibrium the structure of a homogeneous liquid can be quantified by
$G({\bf r})$, the ``density-density'' correlation
function~\cite{stanley,hansen}, defined as $G({\bf r})= \int d{\bf r'}
\Bigl \langle \bigl [ n({\bf r'} + {\bf r}) - \langle n \rangle \bigr
] \, \bigl [ n({\bf r'}) - \langle n \rangle \bigr ] \Bigr \rangle$.
Here, $n({\bf r})=\sum_{i=1}^N \delta \bigl ({\bf r}-{\bf r}_{i} \bigr
)$, and $\langle \dots \rangle$ indicates an ensemble average.
$N=\int d{\bf r} \, n({\bf r})$ is the number of particles present in
a particular configuration. For a homogeneous liquid the density $\rho
= \langle n \rangle=\langle N \rangle/V$.  If the liquid is isotropic,
$G({\bf r})$ further reduces simply to $G(r)$, where $r=|{\bf r}|$.
$G({\bf r})$ measures the spatial correlations of fluctuations of
local density away from the average value.  The pair correlation
function $g({\bf r})$ conventionally presented to characterize the
structure of a liquid is proportional to the ``distinct'' part of
$G({\bf r})$: $G({\bf r}) = \langle N \rangle \delta({\bf r}) +
\langle n \rangle \langle N \rangle \, [g({\bf r}) - 1]$, where
$g({\bf r})$ can be written,
\begin{equation}
g({\bf r}) = {{1}\over{\langle n \rangle \langle N \rangle}}
\Biggl \langle
\sum_{i=1}^N \sum_{{\scriptstyle j=1} \atop {\scriptstyle j\neq i}}^{N}
\delta({\bf r}+{\bf r}_j-{\bf r}_i)
\Biggl \rangle.
\end{equation}
The Fourier transform of $G({\bf r})$ gives the static structure
factor $S({\bf q}) = \Bigl \langle N^{-1} \sum_{i=1}^N \sum_{j=1}^N
\exp\bigr[{-\imath {\bf q}\cdot ({\bf r}_i - {\bf r}_j)}\bigl] \Bigr
\rangle$.

To determine the behavior of $G({\bf r})$ for large $r$, it
is useful to evaluate the fluctuations of $N$, which are related
to the volume integral of $G({\bf r})$, and also related to a
thermodynamic response function, the isothermal compressibility
$\kappa$~\cite{stanley}:
\begin{equation}
\Bigr \langle \bigr [ N - \langle N \rangle \bigl ]^2 \Bigl \rangle \, =
\int d{\bf r} \, G({\bf r}) = \langle n \rangle \langle N \rangle 
kT\kappa, 
\label{fluc}
\end{equation}
where $k$ is Boltzmann's constant.  The convergence or divergence of
the volume integral of $G({\bf r})$ depends on how rapidly $G({\bf
r})$ decays to zero as $r \to \infty$.  If the integral converges,
$G({\bf r})$ is ``short ranged''; if it diverges, $G({\bf r})$ is
``long ranged''.  Near a conventional critical point, $\kappa$
diverges, macroscopic density fluctuations occur, and the behavior of
$G({\bf r})$ approaches that of a long ranged function.

To develop a simple spatial correlation function for a local {\it
dynamical} property in a liquid, we consider for a particle $i$ its
displacement $\mu_i(t,\Delta t)=|{\bf r}_i(t+\Delta t) - {\bf
r}_i(t)|$ over some interval of time $\Delta t$, starting from a
reference time $t$.  We examine the spatial correlations of these
displacements by modifying the definition of $G({\bf r})$ so that the
contribution of a particle $i$ to the correlation function is weighted
by $\mu_i$.  That is, we define a ``displacement-displacement''
correlation function \cite{landau,spinglass},
\begin{eqnarray}
\label{guu}
G_u({\bf r},\Delta t)= \\ \nonumber
\int d{\bf r'} \Bigl\langle \bigl [ u({\bf r'}+{\bf r},t,\Delta t)  
- \langle u \rangle \bigr ] \, \bigl [ u({\bf r'},t,\Delta t)  
- \langle u \rangle \bigr ] \Bigl\rangle,
\end{eqnarray}
where,
\begin{equation}
u({\bf r},t,\Delta t)
=\sum_{i=1}^N 
\mu_i(t,\Delta t)
\, \delta\bigl({\bf r}-{\bf r}_{i}(t)\bigr).
\label{u}
\end{equation}
$G_u({\bf r},\Delta t)$ measures correlations in fluctuations of local
displacements away from their average value.  We are considering an
equilibrium liquid and so $G_u$ does not depend on the choice of the
reference time $t$.  Similarly, $\langle u \rangle \equiv \langle
u({\bf r}, t, \Delta t) \rangle$ does not depend on $t$; for a
homogeneous liquid, it also does not depend on $\bf r$.  In analogy to
the relation between $\langle n \rangle$ and $\langle N \rangle$, we
define the ``total displacement'' $U(t,\Delta t)=\int d{\bf r} \,
u({\bf r},t,\Delta t)$ and its ensemble average $\langle U
\rangle\equiv \langle U(t,\Delta t) \rangle$.  In a constant-$N$
ensemble, both $\langle u \rangle$ and $\langle U \rangle$ are readily
evaluated from the mean displacement ${\overline {\mu}} \equiv \bigl
\langle N^{-1} \sum_{i=1}^N \mu_i(t,\Delta t) \bigr \rangle$: $\langle
u \rangle={\overline \mu} \langle n \rangle$ and $\langle U
\rangle={\overline \mu} \langle N \rangle$.  In equilibrium, $\langle
u \rangle$, $\langle U \rangle$ and ${\overline \mu}$ do not depend on
$t$, but they retain a dependence on $\Delta t$.

$G_u({\bf r},\Delta t)$ can be written so as to identify a spatial
correlation function $g_u({\bf r},\Delta t)$ analogous to $g({\bf
r})$:
\begin{equation}
G_u({\bf r},\Delta t) = {\bigl \langle N \bigr \rangle} {\overline
{\mu^2}} \, \delta({\bf r}) + \langle u \rangle \langle U \rangle \,
[g_u({\bf r},\Delta t) - 1],
\end{equation}
where
\begin{eqnarray}
g_u({\bf r},\Delta t) =
{{1}\over{
\langle u \rangle \langle U \rangle 
}} \\ \nonumber
\times
\Biggl \langle
\sum_{i=1}^N \sum_{{\scriptstyle j=1} \atop {\scriptstyle j\neq i}}^{N}
\mu_i(t,\Delta t) \, \mu_j(t,\Delta t) \,
\delta\bigl({\bf r}+{\bf r}_j(t)-{\bf r}_i(t)\bigr)
\Biggl \rangle. 
\end{eqnarray}
The mean squared displacement ${\overline {\mu^2}} \equiv \bigl
\langle N^{-1} \sum_{i=1}^N \mu_i^2(t,\Delta t) \bigr \rangle$, and
also depends on $\Delta t$.  The Fourier transform of $G_u({\bf
r},\Delta t)$ gives a ``structure factor'' $S_u({\bf q},\Delta t)$ for
the particle displacements: $S_u({\bf q},\Delta t) = \Bigl \langle (N
{\overline \mu}^2)^{-1} \sum_{i=1}^N \sum_{j=1}^N \mu_i(t,\Delta t)
\mu_j(t,\Delta t) \exp\bigl [{-\imath {\bf q}\cdot \bigl ({\bf r}_i(t)
- {\bf r}_j(t)\bigr )}\bigr ] \Bigr \rangle$.

In analogy to Eq.~\ref{fluc}, the fluctuations of $U$ are related to
the volume integral of $G_u({\bf r},\Delta t)$:
\begin{equation}
\Bigl \langle \bigl [ U - \langle U \rangle \bigr ]^2
\Bigr \rangle = \int d{\bf r} \, G_u({\bf r},\Delta t) \equiv 
\langle u \rangle \langle U \rangle kT \kappa_u.
\label{flucuu}
\end{equation}
We have defined the quantity $\kappa_u$ in analogy to $\kappa$.
Hence, as for $G({\bf r})$, we can determine the large $r$ behavior of
$G_u({\bf r},\Delta t)$ from the fluctuations of a bulk quantity, $U$.

To evaluate these quantities we use data obtained
\cite{kdppg,ddkppg,dgpkp} from a molecular dynamics simulation of a
model Lennard-Jones glass-former.  The system is a three-dimensional
binary mixture (80:20) of 8000 particles interacting via Lennard-Jones
interaction parameters \cite{units}.  We analyze data from seven
$(\rho,P,T)$ state points on a line in the $P,T$ plane approaching
$T_c \approx 0.435$ at a pressure $P \approx 3.03$~\cite{kob}. (In the
remainder of this paper, all values are quoted in reduced units
\cite{units}).  The highest and lowest $T$ state points simulated are
$(\rho = 1.09, P=0.50, T=0.550)$ and
$(\rho=1.19,P=2.68,T=0.451)$. Following equilibration at each state
point, the particle trajectories are monitored in the $NVE$ ensemble
($E$ is the total energy) for up to $1.2 \times 10^4$ Lennard-Jones
time units (25.4 ns in argon units) for the coldest $T$. Complete
simulation details may be found in \cite{dgpkp}.  All quantities
presented here are calculated using all 8000 particles in the
liquid. The results presented here do not change when the minority
particles are excluded \cite{sametc}.

For all seven state points, a ``plateau'' exists in both ${\overline
{\mu^2}}$ and the self part of the intermediate scattering function
$F_s(q,t)$ as a function of $t$ \cite{dgpkp}.  The plateau separates
an early time ballistic regime from a late time diffusive regime, and
indicates ``caging'' of the particles typical of low $T$, high $\rho$
liquids.  The $\alpha$-relaxation time $\tau_{\alpha}$ describes the
decay of $F_s(q,t)$ to zero at the value of $q$ corresponding to the
first peak in the static structure factor $S(q)$.  Over the range of
$T$ studied, $\tau_{\alpha}$ increases by 2.4 orders of magnitude, and
follows a power law $\tau_{\alpha} \sim (T-T_c)^{-\gamma}$, with $T_c
= 0.435$ and $\gamma \simeq 2.8$.  The diffusion coefficient $D$
follows a power law $D \sim (T-T_c)^{\gamma}$, with $T_c = 0.435$ and
$\gamma \simeq 2.13$, and thus diffusion and structural relaxation are
``decoupled'' \cite{decoupled}.  The simulated liquid states analyzed
here therefore exhibit the complex bulk relaxation behavior
characteristic of a supercooled liquid approaching its glass
transition.  Both $g(r)$ and $S(q)$ for this liquid have been
calculated previously \cite{dgpkp,kob}, and it has been shown that as
$T$ decreases, no long range structural correlations due to density or
composition fluctuations occur.

\begin{figure}
\hbox to\hsize{\epsfxsize=1.0\hsize\hfil\epsfbox{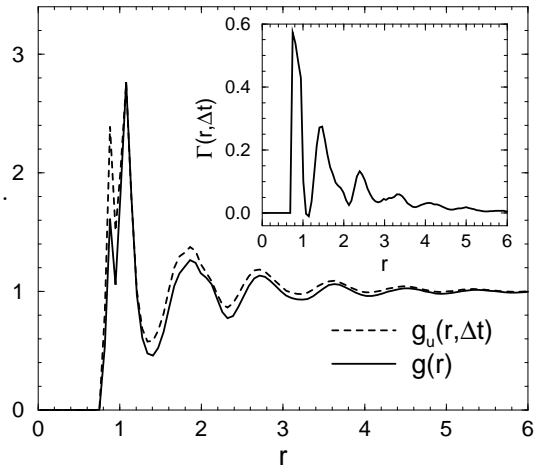}\hfil}
\caption{$g_u(r,\Delta t)$ and $g(r)$ versus $r$ at $T=0.451$. $\Delta
t$ is chosen on the order of $\tau_{\alpha}$. Inset: $\Gamma(r,\Delta
t)$ versus $r$.}
\label{figguu}
\end{figure}

In Fig.\ref{figguu} we show $g_u(r,\Delta t)$ as a function of $r$ for
$T=0.451$, and with $\Delta t$ chosen to be on the order of
$\tau_{\alpha}$.  $g(r)$ for the same $T$ is also shown.  For a fixed
choice of $\Delta t$, note that if the displacement were always the
same for every particle, then $g_u(r,\Delta t)$ and $g(r)$ would be
identical for all $r$.  Hence, it is deviations of $g_u(r,\Delta t)$
from $g(r)$ that will inform us of spatial displacement correlations
in excess of those that would be expected based on a knowledge of
$g(r)$ alone.  We find that for this choice of $\Delta t$,
$g_u(r,\Delta t)$ is appreciably higher than $g(r)$ for values of $r$
up to several interparticle distances. This excess correlation is made
clearer in the inset of Fig.\ref{figguu}, where we show the function
$\Gamma(r,\Delta t)\equiv [g_u(r,\Delta t)/g(r)]-1$.

\begin{figure}
\hbox to\hsize{\epsfxsize=1.0\hsize\hfil\epsfbox{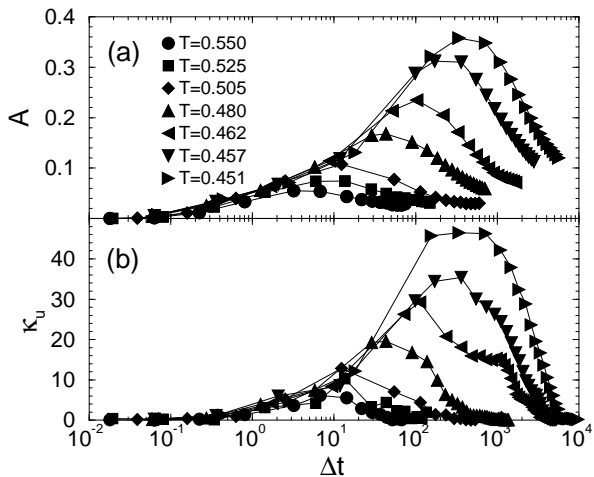}\hfil}
\caption{(a) $A$ versus $\Delta t$ for different $T$.  (b) $\kappa_u$
as a function of $\Delta t$ for the same $T$ as in (a).}
\label{figAdt}
\end{figure}

\begin{figure}
\hbox to\hsize{\epsfxsize=1.0\hsize\hfil\epsfbox{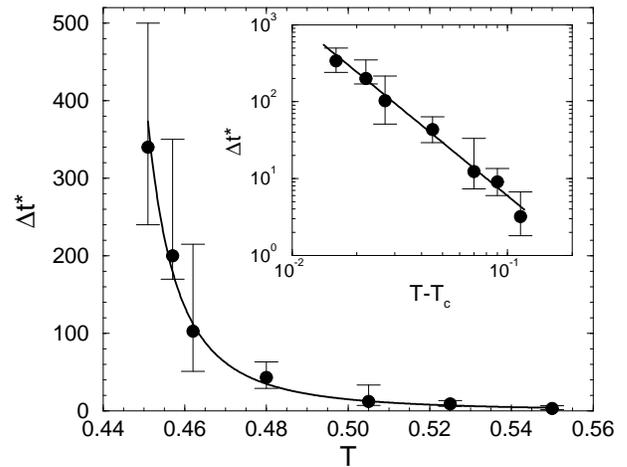}\hfil}
\caption{$\Delta t^*$ plotted versus $T$. The solid curve is a
power-law fit to the data. INSET: Log-log plot of $\Delta t^*$ versus
$T-T_c$, and the power-law fit to the data. $T_c=0.435$.}
\label{figdt}
\end{figure}

However, the question arises as to how to select the value of $\Delta
t$.  We find that the behavior of the liquid itself suggests a unique
choice for $\Delta t$.  To demonstrate this, we show in
Fig.~\ref{figAdt} the total excess correlation $A \equiv \int dr \,
\Gamma(r,\Delta t)$ as a function of $\Delta t$.  We find that there
is a value of $\Delta t=\Delta t^*$ at which $A$ is a maximum and that
both the maximum value of $A$ and $\Delta t^*$ increase with
decreasing $T$.  Hence for each $T$ the spatial correlation of
particle displacements is most prominent at $\Delta t^*$.  Moreover,
all curves for $T \le 0.525$ collapse onto a single master curve when
$t$ is scaled by $\Delta t^*$ and $A$ is scaled by $A(\Delta t^*)$,
suggesting that $\Delta t^*$ is a characteristic time for this
liquid. In the remainder of this Letter, all quantities are therefore
evaluated for $\Delta t=\Delta t^*$.  Fig.~\ref{figdt} shows that
$\Delta t^*$ follows a power law with $T$: an excellent fit~\cite{fit}
to the form $\Delta t^* \sim (T-T_c)^{-\gamma}$ is obtained when $T_c
= 0.435$, and yields $\gamma = 2.3 \pm 0.2$. This value for $\gamma$
is different from the exponent found for $\tau_{\alpha}$, but (within
our numerical uncertainty) cannot be distinguished from the exponent
governing the apparent vanishing of $D$ at $T_c$. 

If $A$ is largest at $\Delta t^*$, then we might also expect $\kappa_u$
to be largest at $\Delta t^*$, since by Eq.~\ref{flucuu} $\kappa_u$
quantifies the total magnitude (integrated over space) of the
displacement correlations quantified by $G_u({\bf r},\Delta t)$.  We
evaluate $\kappa_u$ from the fluctuations of $U$ according to
Eq.~\ref{flucuu} (Fig.~\ref{figAdt}b) and confirm $\kappa_u$ exhibits
the same behavior as $A$: $\kappa_u$ goes to zero at short and long
times, and has a maximum at a $T$-dependent $\Delta t^*$.

\begin{figure}
\hbox
to\hsize{\epsfxsize=1.0\hsize\hfil\epsfbox{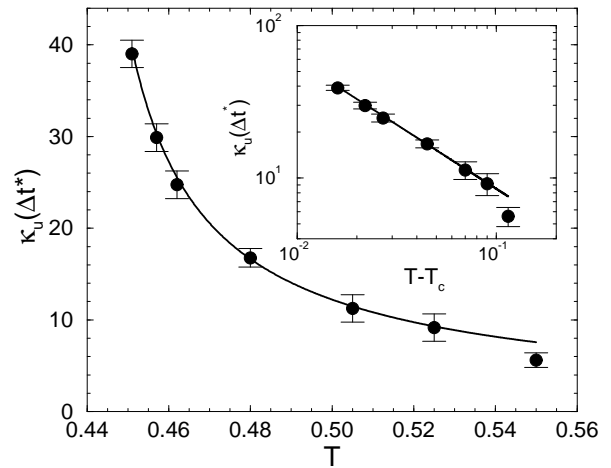}\hfil}
\caption{$\kappa_u(\Delta t^*)$ plotted versus $T$. The solid curve is
a power-law fit to the data.  INSET: Log-log plot of $\kappa_u(\Delta
t^*)$ versus $T-T_c$, and the power-law fit to the data. $T_c=0.435$.}
\label{figchiT}
\end{figure}

In Fig.~\ref{figchiT} we show the $T$-dependence of $\kappa_u$ for
$\Delta t=\Delta t^*$. We find that $\kappa_u(\Delta t^*)$ grows
monotonically with decreasing $T$, indicating that the range of the
correlation measured by $G_u(r,\Delta t^*)$ is growing with decreasing
$T$. We find that a power law $\kappa_u(\Delta t^*) \sim
(T-T_c)^{-\gamma}$ fits well to the data when $T_c = 0.435$, and gives
$\gamma = 0.84$.  Thus $\kappa_u$ exhibits an apparent divergence at a
$T$ that is within numerical error of $T_c$, demonstrating that
$G_u(r,\Delta t^*)$ is becoming increasingly long-ranged as $T \to
T_c$.

\begin{figure}
\hbox to\hsize{\epsfxsize=1.0\hsize\hfil\epsfbox{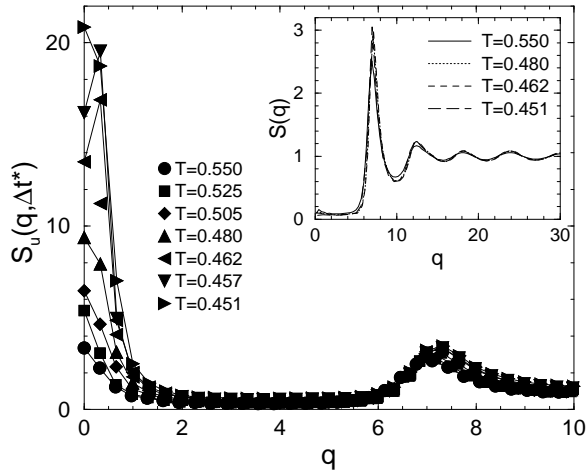}\hfil}
\caption{$S_u(q,\Delta t^*)$ versus $q$ for different $T$. The values
at $q=0$ are obtained from the fluctuations in $U$ via the relation
$S_u(q=0,\Delta t^*) = \rho T \kappa_u(\Delta t^*)$.  INSET: Static
structure factor $S(q)$ for four different $T$.}
\label{figsuu}
\end{figure}

To estimate a correlation length associated with these displacement
correlations, we evaluate $S_u( q,\Delta t^*)$ for different $T$
(Fig.~\ref{figsuu}). For intermediate and large $q$, $S_u(q,\Delta
t^*)$ coincides with $S(q)$.  However, for $q\rightarrow 0$ a peak
develops and grows with decreasing $T$, again demonstrating the
presence of long range dynamical correlations.  No growing peak at
$q=0$ appears in the static structure factor $S(q)$
(Fig.~\ref{figsuu}, inset).  To quantify the correlation length, we
attempted to fit $S_u(q,\Delta t^*)$ using an Orstein-Zernike form,
$S_u(q)\propto 1/(1+\xi^2 q^2)$, where $\xi$ is the correlation
length.  Although this form fits well to the data at the highest $T$,
it fails at lower $T$, making the interpretation of the fitted $\xi$
values ambiguous.

Nevertheless, it was shown previously for this system that highly
``mobile'' particles move cooperatively \cite{ddkppg} and form
clusters \cite{kdppg} whose mean size diverges at $T_c$ \cite{dgpkp}.
These clusters contribute to the growing range of $G_u(r,\Delta t^*)$,
and thus they can be used to give a rough estimate of the length scale
over which particle motions are correlated. At $T=0.451$, this average
length scale exceeds 3 particle diameters, and the largest cluster has
a length scale that exceeds the size of our simulation box
(approximately 19 particle diameters on a side.)

In summary, we have defined a correlation function that quantifies the
spatial correlation of single-particle displacements in a liquid.
Using this function, we have shown in computer simulations of an
equilibrium liquid that the displacements of particles are spatially
correlated over a range and time scale that both grow with decreasing
$T$ as the mode coupling temperature is approached.  While MCT makes
no predictions concerning a growing dynamical correlation
length~\cite{arndt,mountain}, calculation of the vector
displacement-displacement correlation function may be tractable within
the mode coupling framework.  We have also identified a bulk dynamical
variable $U$ whose fluctuations appear to diverge at $T_c$. Hence, $U$
is behaving much like a static order parameter on approaching a
second-order phase transition.  Our analysis therefore suggests that
an extension to dynamically-defined quantities of the framework of
ordinary critical phenomena may be useful for understanding the nature
of supercooled, glass-forming liquiods.

We are grateful to J. Baschnagel, J.F. Douglas, and R. D. Mountain for
helpful feedback.  PHP acknowledges the support of NSERC (Canada).


\begin{thebibliography}{999}
\bibitem{units} The Lennard-Jones interaction parameters
$\epsilon_{\alpha,\beta}$ and $\sigma_{\alpha,\beta}$ are given by:
$\epsilon_{AA}=1.0$, $\epsilon_{AB}=1.5$, $\epsilon_{BB}=0.5$,
$\sigma_{AA}=1.0$, $\sigma_{AB}=0.8$, $\sigma_{BB}=0.88$. Lengths are
defined in units of $\sigma_{AA}$, temperature $T$ in units of
$\epsilon_{AA}$, and time $t$ in units of
$\sqrt{\sigma_{AA}^2m/\epsilon_{AA}}$. Both types of particles have
the same mass, $m$.
\bibitem{sametc} For this same system simulated along a different
path, the same value of $T_c$ was found for the majority particles as
for the minority particles \cite{kob}.
\bibitem{fit} All power law fits presented here were obtained by performing
a linear regression on the log of the data.
\end{thebibliography}
\end{document}